\documentclass[journal]{IEEEtran}

\usepackage[hidelinks]{hyperref}
\usepackage{amsmath,nccmath}
\usepackage{graphicx}
\usepackage{float}
\usepackage{cite}
\usepackage{verbatim}
\usepackage[ruled,vlined]{algorithm2e}

\begin{document}
\setlength{\abovedisplayskip}{1pt}
\setlength{\belowdisplayskip}{1pt}
\setlength{\abovecaptionskip}{-3pt}

\title{A Two-Stage Optimal Bidding Algorithm for Incentive-based Aggregation of Electric Vehicles in Workplace Parking Lots}

\author{Zhongyang~Zhao,~\IEEEmembership{Student~Member,~IEEE,}
        and~Caisheng~Wang,~\IEEEmembership{Senior~Member,~IEEE}\vspace{-6ex}
    
\thanks{The authors are with the Department
of Electrical and Computer Engineering, Wayne State University, Detroit,
MI, 48202 USA (e-mail: zhongyang.zhao@wayne.edu, cwang@wayne.edu). This work was supported in part by the National Science Foundation of USA under Grant IIS-1724227.}
}


%

\maketitle

\begin{abstract}
This paper proposes an incentive-based aggregator for electric vehicles (EVs) in workplace parking lots to participate in the energy and regulation markets. With the implementation of seasonal Autoregressive Integrated Moving Average (ARIMA) model to predict the market information, a Day-Ahead (DA) planning model coordinated with the EV's responses to the incentive is formulated to determine the days to activate the aggregation program and the optimal incentives that are broadcasted to the EV owners in the first stage. Given the determined incentives and EV's responses, an real-time (RT) optimal bidding algorithm complying with EVs' energy demand in the second stage is designed to maximize the aggregator's profits in the markets. The proposed models are tested using the data collected from PJM's energy and regulation markets. The results show the incentive-based aggregator can benefit both the EV owners and the aggregator from the credits obtained from the markets. Meanwhile, the results also indicate the proposed optimal bidding algorithm is capable of handling the uncertainty of regulation signals and following the signals with high precision.
\end{abstract}


%
\IEEEpeerreviewmaketitle

\vspace{-0.4cm}
\section{Introduction} \label{introduction}
\vspace{-0.1cm}
As a type of fast responding and flexible load, electric vehicles (EVs) have been investigated to provide system-wide services and participate in electricity markets \cite{PJMEVtest,ortega2012electric,bessa2010role}. It has been shown that EVs are capable of following the regulation signals from PJM \cite{PJMEVtest}. While a single EV with a limited capacity is not satisfied to participate in the wholesale markets, an EV aggregator, as an essential role between EV owners and the wholesale markets, can be implemented to integrate and coordinate a fleet of EVs to enter the markets \cite{bessa2010role}. Since the global sales of EVs reached 2.1 million in 2018 \cite{2018EVVolumes}, the large amount of EVs has challenged the management of EV aggregation. The EV aggregator, as a third-party service, is commissioned to provide the EV owners benefits and operate in competitive wholesale markets profitably. One of the key implementations is to aggregate the EVs at public parking lots \cite{seddig2019two, neyestani2014allocation}. The peak hours of market conditions in both energy and regulation markets, such as locational marginal price (LMP) and regulation market-clearing prices, are between 7:00 and 21:00 \cite{zhao2017improving, zhao2020revenue, donadee2014agc}, which largely overlap with many people's working schedule. Hence, the aggregator of workplace parking lots has a great potential to compete in the wholesale market during the peak hours of electricity prices. \par
Compared to the other demand response methods \cite{shafie2016optimal,nguyen2018bidding}, such as the real-time pricing method, the incentive-based methods have the advantages of allowing the customers to participate voluntarily. Furthermore, the incentive information is not required to be always broadcasted to the customers. In other words, the incentive-based EV aggregation program can be activated only when it is beneficial for both EV aggregator and owners. Meanwhile, the incentive is able to not only compensate the EV owners but also attract more EVs to join the aggregation program. The incentive-based demand response programs introduced in \cite{wang2016two,zhong2013coupon,fang2016coupon,do2018stochastic,sarker2014optimal} have proven the ability to improve the demand flexibility in retail customers. In addition, according to the revenue analysis of battery in energy and regulation markets in \cite{zhao2020revenue, byrne2016estimating}, the revenue obtained from the regulation market far exceeds that from the energy market. Most importantly, it can raise more regulation resources for power system operations. In this paper, an EV aggregator is proposed to participate in both the energy and regulation markets for a greater potential of revenue. \par
However, there are still highly challenging issues, such as the uncertainties of EV owners' behaviors and market conditions, which can derail the aggregator's performance. The EV owners’ behaviors, including their arrival/departure times, the EV batteries’ State of Charge (SOC) at arrival/departure, and their responses to the aggregator's control signals, expose the EV aggregator to a complicated and risky situation. To address these challenges, several studies have been carried out for helping EV aggregators to more effectively participate in the wholesale markets \cite{bessa2012optimized,vagropoulos2013optimal,sarker2015optimal,shafie2016optimal,yao2016optimization,liu2018two,vaya2014optimal}. An optimal bidding strategy based on the two-stage stochastic linear programming was developed in \cite{vagropoulos2013optimal} for an EV aggregator to participate in DA energy and regulation markets with considering the uncertainties of market data and EV fleet. In \cite{shafie2016optimal}, an EV aggregator model was proposed to study the operational behaviors of a parking lot with different demand response programs by changing the energy demand of EVs. In \cite{yao2016optimization}, a stochastic optimization model considering the conditional value at risk was proposed for the EV aggregator to participate in the frequency regulation market. Based on the transactive control, a two-stage optimal charging scheme was implemented in \cite{liu2018two} to minimize the total cost of EV aggregator by managing the energy procurement in the DA market and scheduling EV charging. An optimal bidding strategy considering each individual vehicle's behaviors was proposed in \cite{vaya2014optimal} to bid in the DA market for minimizing the charging cost of the aggregator. \par
Nevertheless, the effectiveness and implementation of incentives to induce the EV owners to adjust their behaviors for providing more resources for the EV aggregator to participate in the wholesale markets have rarely been studied. In this paper, a two-stage optimal bidding algorithm for an aggregator of EVs in workplace parking lots to participate in RT energy and regulation markets, coordinated with the incentive to adjust the EVs’ behaviors, is designed to benefit both EV aggregator and EV owners. 
The main contributions of this work can be summarized as follows: 1) To secure more resources for the aggregator when needed and address the uncertainty of EV owners' behaviors, the EV's responses to the incentives are modeled and considered in the proposed two-stage optimal bidding algorithm for the EV aggregator to participate in both RT energy and regulation markets. 2) In the first stage, a DA planning model based on the predictions of the seasonal ARIMA model is proposed for the EV aggregator to determine if the EV aggregation program should be activated for the next day and what level of the incentive should be broadcasted to optimally change the EVs' behaviors. 3) Based on the incentive determined in the previous stage, a RT optimal bidding strategy is designed for the EV aggregator to maximize the profit in the RT energy and regulation markets, complying with the energy requirements of EVs. \par
%
%
The remainder of this paper is organized as follows: Section \ref{Sec_EV_Behaviors} introduces the modeling of EV owners’ behaviors and responses to incentives. In Section \ref{twostage_optimal_bidding}, a two-stage optimal bidding algorithm for the incentive-based EV aggregator is proposed. Section \ref{model_performance} tests the performances of the proposed models with the market data collected from PJM. The conclusion is drawn in Section \ref{conclusion}. \par \vspace{-0.3cm}

\section{Modeling of EV Owners' Behaviors} \label{Sec_EV_Behaviors}
\vspace{-0.05cm}
%
%
The EV aggregator in this study is designed to provide two types of incentives to the EVs. The first one is a fixed reward $\hat{\pi}$. Once the EV aggregation program is activated, there is a fixed reward provided for each EV owner. This reward is designed for attracting EV owners to participate in the EV aggregation program. The other one is the incentive $\pi$ designed to attract the EV owners to change their behaviors for providing more resources for the aggregator to participate in the energy and regulation markets when needed. \par
In order to study the responses of an individual EV owner to different incentives, the truncated Gaussian distributions are used to model the original EV behaviors, including the arrival time, departure time, SOC when arriving, and desired SOC when leaving. For example, the distribution of arrival time can be modeled by using a truncated Gaussian distribution ($f_{TG}$): \vspace{-0.4cm}
\begin{equation}\label{CIEV_eq1} t_{arv}=f_{TG}(x;{\mu }_{t_{arv}},{\sigma }_{t_{arv}},t^{min}_{arv},t^{max}_{arv}) \vspace{-0.2cm} \end{equation}
Based on \cite{shafie2016optimal,bessa2012optimized,vagropoulos2013optimal}, the parameters of the distributions for modeling the EV owners’ behaviors are listed in Table \ref{tb_Para_Orgin_Behavior}. For the arrival and departure times, the mean values indicate that people would mostly like to go to work at around 8:30 AM and leave at around 5:30 PM; min values show the earliest time when people arrive/depart and max values represent the latest time of arrival/departure. Similarly, the parameters of SOC at arrival/departure represent the randomness of the EV owners’ behaviors in charging their EVs. Given the parameters listed in Table \ref{tb_Para_Orgin_Behavior}, the initial behaviors of each EV owner can be modeled by the corresponding truncated Gaussian distributions. Since the EV aggregator is proposed to participate in the RT markets on the hourly basis, the EV arriving between 8:00 and 8:59 is considered as arriving within the hour of 8.
\begin{table}
	\centering
	\caption{Parameters of EV Original Behaviors' Distributions}\label{tb_Para_Orgin_Behavior}
	\begin{tabular}{c|cccc} \hline 
		& \textbf{Mean} & \textbf{Std.} & \textbf{Min} & \textbf{Max} \\ \hline 
		\textbf{Arrival Time (h)} & 8.5 & 3 & 6 & 13 \\ 
		\textbf{Departure Time (h)} & 17.5 & 3 & 13 & 20 \\ 
		\textbf{SOC at Arrival (\%)} & 75 & 25 & 25 & 95 \\ 
		\textbf{SOC at Departure (\%)} & 90 & 10 & 60 & 100 \\ \hline 
	\end{tabular}
\vspace{-0.7cm}
\end{table} \par
%
With the initial behaviors of each EV owner, the step function for modeling the EV owner's response to different levels of incentives can be built. As formulated in \eqref{eq_Step_tev_arv}, the arrival time $t_{arv}^i$ of EV $i$ with different incentives are described with three intervals. It shows the EV $i$ initially arrives at hour 9 and is willing to come at hour 7 when the incentive $\pi$ is over $\pi_2^i$. Similarly, the step response functions of departure time $t_{dep}^i$, arrival SOC $SOC_{arv}^i$ and departure SOC $SOC_{dep}^i$ for EV $i$ can be developed. In these step functions, the minimum and maximum values of the different responses are the same as the initial behavioral parameters as shown in Table \ref{tb_Para_Orgin_Behavior}. The advantage of estimating the response behavior individually is the EV owners can be characterized with the diversity of response elasticity by different step functions.
\begin{equation}\label{eq_Step_tev_arv}
t_{arv}^i = \begin{cases}
9, & 0 \leq \pi < \pi_{1}^i \\
8, &  \pi_{1}^i \leq \pi < \pi_{2}^i \\
7, & \pi_{2}^i \leq \pi < \pi_{3}^i.
\end{cases}
\end{equation}
Given the step functions of each EV’s responses, the individual responses can be aggregated to the responses of the whole EV fleet with different incentives. For example, the number ($N_t^{EV,arv}$) of newly arriving EV available to the aggregator at hour $t$ with different incentives can be obtained as \eqref{eq_Step_Nev_arv}:
\begin{equation} \label{eq_Step_Nev_arv}
{N}_{t}^{EV,arv} = \begin{cases}
{N}_{t,1}^{EV,arv}, & 0 \leq \pi < \pi_{1} \\
\vspace{-0.1cm}
\vdots &  \vdots \\
\vspace{-0.1cm}
{N}_{t,\omega}^{EV,arv}, & \pi_{\omega-1} \leq \pi < \pi_{\omega} \\
\vspace{-0.1cm}
\vdots &  \vdots \\

{N}_{t,\Omega}^{EV,arv}, & \pi_{\Omega-1} \leq \pi < \pi_{\Omega} \\
\end{cases}
\end{equation}
where $\omega$ indicates the $\omega$-th interval of the incentive $\pi$ and $\Omega$ is the number of all intervals in the response functions of the whole EV fleet; $N_{t,\omega}^{EV,arv}$ means the number of EV newly available for the EV aggregator at hour $t$ if the incentive $\pi$ is between $\pi_{\omega-1}$ and $\pi_{\omega}$. It should be noted that $t_{arv}^i$ indicates EV $i$ arrives at the parking lot within that hour, which means if the EVs arrive at hour $t$, they are not available for EV aggregator to operate until hour $t+1$. Hence, the first hour for $N_t^{EV,arv}$ at the aggregator is $t=7$ while the earliest arrival time of EVs is hour 6. Similarly, the number of EV leaving parking lot $N_t^{EV,dep}$ can be obtained and the last operation hour for the aggregator to participate in the markets is set as $t=19$ for the EV owners who leave the parking lot at hour 20. In addition to $ N_t^{EV,arv}$  and $ N_t^{EV,dep}$, the arrival energy $E_t^{EV,arv}$ and desired departure energy $E_t^{EV,dep}$ of the EV fleet at hour $t$ with different incentives can be derived from individual SOC response functions and EV capacity $E_{max}^{EV}$. 

\vspace{-0.2cm}
\section{Two-Stage Optimal Bidding Algorithm} \label{twostage_optimal_bidding}
\vspace{-0.0cm}
\subsection{First Stage: Day-Ahead Planning of EV Aggregator}
\vspace{-0.1cm}
As aforementioned, the EV aggregator is designed to provide two types of incentives to the EV owners. The incentive $\pi$ is for inducing the EV owners to change their behaviors. The incentive information is required to be sent to the EVs one day ahead in order to change the EV behaviors of the next day. Therefore, a DA planning model considering the market information is important for the EV aggregator to determine the optimal incentive signal $\pi$ delivered to the EV owners. \par
In the following DA planning model, the objective function \eqref{DA_DAobj}, subject to \eqref{DA_sumW}-\eqref{DA_restS}, is to maximize the profit for EV aggregator to participate in the energy and regulation markets, which are represented by $Credit_{E,\omega}$ and $Credit_{R,\omega}$, with distributing the fixed reward $\hat{\pi}$ and the incentive $\pi$ to EV owners. $W_{\omega}$ is a binary weight related to the intervals of $\pi$. \eqref{DA_sumW} enforces the DA planning model to maximize the profit within only one interval of $\pi$. Each interval of $\pi$ is defined as \eqref{DA_PiInterval}. For example, $W_1 = 1$ and $W_\omega=0$ for $\omega\neq 1$ when $\pi_{0} \leq \pi < \pi_1$.
\begin{gather}
Max \sum^{\Omega}_{\omega = 1} W_{\omega} (Credit_{E,\omega}+Credit_{R,\omega}-(\hat{\pi}+\pi)) \label{DA_DAobj}
\end{gather}
\begin{center}\vspace{-0.2cm} s.t. \eqref{DA_sumW}-\eqref{DA_restS}. \vspace{-0.1cm}\end{center}
\begin{gather}
\sum^{\Omega}_{\omega = 1} W_{\omega} = 1 \label{DA_sumW}\\
W_{\omega} \in \{0,1\} \\
\pi_{\omega-1} \leq \pi < \pi_\omega \label{DA_PiInterval}
\end{gather} \par
\textbf{Credits in Markets}:
When $\pi \in [\pi_{\omega-1},\pi_{\omega})$, the credit of energy market $Credit_{E,\omega}$ and the credit of regulation market $Credit_{R,\omega}$ are determined by \eqref{DA_CreditE} and \eqref{DA_CreditR}, where $P_{t,\omega}^{E,ch}$, $P_{t,\omega}^{E,dis}$, and $P_{t,\omega}^{R}$ respectively represent the charging power, discharging power, and regulation power offered in the markets. These power offers are the decision variables determined by the DA planning model to maximize the profit of the EV aggregator based on the estimated market information, including the LMP, $\overline{LMP}_t$, Regulation Market Capability Clearing Price, $\overline{RMCCP}_t$, Regulation Market Performance Clearing Price, $\overline{RMPCP}_t$, performance score $\overline{\rho}_t$, and mileage ratio $\overline{\beta}_t$, estimated for the next day (Note: The method to estimate the market information for the next day is introduced in Section \ref{Sec_SARIMA}). The performance score $\overline{\rho}_t$ is used to evaluate how well the regulation resources follow the regulation signal and penalize the resources that fail to follow. In the PJM's regulation market, there are two types of regulation signals: Regulation A and Regulation D, ranging between 0 and 1. Compared to the fast and dynamic Regulation D signal, Regulation A is a slower signal. Considering the battery is capable of fast responding to the dispatch signals, the EV aggregator is proposed to follow Regulation D. The mileage ratio in the regulation market measures the mileages of Regulation D over Regulation A, which is used to evaluate the relative movement of following the different two signals. For further clarifications of the market data, please refer to the PJM Manual \cite{PJMMaunual11} and related studies \cite{zhao2020revenue, byrne2016estimating, xu2018optimal}.
\begin{gather} \allowdisplaybreaks
Credit_{E,\omega} = \sum^{}_{t \in T} (P_{t,\omega}^{E,dis}-P_{t,\omega}^{E,ch})\times \overline{LMP}_t \label{DA_CreditE}\\
\begin{split}
Credit_{R,\omega} = & \sum^{}_{t \in T} (P^R_{t,\omega}\times \overline{{RMCCP}}_t \times\ \overline{{\rho}}_t + \\ 
&P^R_{t,\omega}\times \overline{{RMPCP}}_t \times \overline{\beta}_t \times\ \overline{{\rho}}_t)  \label{DA_CreditR}
\end{split} 
\end{gather} \par
\textbf{Power Constraints}:
By aggregating the $N_{t,\omega}^{EV,arv}$ and $N_{t,\omega}^{EV,dep}$ from the first operation hour to hour $t$, the number of EVs at the aggregator $ N_{t,\omega}^{EV }$ can be obtained, and the max power of the aggregator $P_{max,t,\omega}^{Agg}$ at hour $t$ can be calculated by \eqref{DA_maxPagg}. \eqref{DA_maxPEch}-\eqref{DA_maxPR} enforces the offers in the energy and regulation markets could not exceed the maximum power of each hour. Since the aggregator cannot charge and discharge in the energy market at the same time, a binary decision variable $\delta_{t,\omega}$ is implemented in \eqref{DA_bigMch} and \eqref{DA_bidMdis} to enforce the aggregator to only offer either charging or discharging power in a given hour segment. Meanwhile, the total power offered to the energy and regulation markets cannot exceed the maximum power of the aggregator at hour $t$ as described by \eqref{DA_maxTotalPowerch} and \eqref{DA_maxTotalPowerdis}. Given the $P_{t,\omega}^{E,ch}$, $P_{t,\omega}^{E,dis}$, and $P_{t,\omega}^{R}$, the total charging power $P_{t,\omega}^{Agg,ch}$ and discharging power $P_{t,\omega}^{Agg,dis}$ of the aggregator can be obtained by \eqref{DA_Paggch} and \eqref{DA_Paggdis}, where the $\overline{RegD}_t^{down}$ and $\overline{RegD}_t^{up}$ are the estimated Regulation D sent by the Independent System Operator (ISO) or Regional Transmission Organization (RTO) to conduct the charge and discharge outputs in the regulation market, respectively. \par
\begin{gather}
P_{max,t,\omega}^{Agg} = {N}_{t,\omega}^{EV} \times P_{max}^{EV} \label{DA_maxPagg}\\
0 \leq P_{t,\omega}^{E,ch} \leq P_{max,t,\omega}^{Agg} \label{DA_maxPEch}\\
0 \leq P_{t,\omega}^{E,dis} \leq P_{max,t,\omega}^{Agg} \label{DA_maxPEdis}\\
0 \leq P_{t,\omega}^{R} \leq P_{max,t,\omega}^{Agg} \label{DA_maxPR}\\
0 \leq P_{t,\omega}^{E,ch} \leq \delta_{t,\omega} \times M \label{DA_bigMch}\\
0 \leq P_{t,\omega}^{E,dis} \leq (1-\delta_{t,\omega}) \times M \label{DA_bidMdis}\\
\delta_{t,\omega}  \in \{0,1\} \\
0 \leq P_{t,\omega}^{E,ch}+P_{t,\omega}^{R} \leq P_{max,t,\omega}^{Agg} \label{DA_maxTotalPowerch}\\
0 \leq P_{t,\omega}^{E,dis}+P_{t,\omega}^{R} \leq P_{max,t,\omega}^{Agg} \label{DA_maxTotalPowerdis}\\
P_{t,\omega}^{Agg,ch} = P_{t,\omega}^{E,ch}+P_{t,\omega}^R \times \overline{RegD}_t^{down} \label{DA_Paggch}\\
P_{t,\omega}^{Agg,dis} = P_{t,\omega}^{E,dis}+P_{t,\omega}^R \times \overline{RegD}_t^{up} \label{DA_Paggdis}
\end{gather} \par
\textbf{Energy Constraints}:
In addition to the constraints \eqref{DA_maxPagg}-\eqref{DA_Paggdis} on the power outputs of the aggregator, there are also constraints on the EV aggregator's energy amount in the DA planning model. \eqref{DA_St} specifies the aggregator's energy $E_{t,\omega}^{Agg}$ at hour $t$, which is obtained according to the aggregator's energy $E_{t-1,\omega}^{Agg}$ of the previous hour, the aggregator's departure energy $E_{t, \omega}^{Agg,dep}$,  the energy $E_{t,\omega}^{EV,arv}$ provided by the new arrival EVs, and the total aggregated output at hour $t$. In \eqref{DA_St}, the $\eta_c$ and $\eta_d$ are the charging and discharging efficiencies, respectively. \eqref{DA_Saggdep} shows the $E_{t,\omega}^{Agg,dep}$ should be not only larger than desired departure energy $E_{t,\omega}^{EV,dep}$ of the EVs, which is obtained from response functions, but also smaller than maximum energy of the departure EVs obtained by the number $N_{t,\omega}^{EV,dep}$ of EVs leaving the parking lot at hour $t$ and the capacity $E_{max}^{EV}$ of each EV. In other words, the aggregator can leave more departure energy than the demand of EVs when they depart. Therefore, in order to make the aggregators' operation more flexible, the $E_{t,\omega}^{Agg,dep}$ is implemented in \eqref{DA_St} instead of $E_{t,\omega}^{EV,dep}$. As shown in \eqref{DA_StLimit}, the $E_{t,\omega}^{Agg}$ should be able to fulfill the energy demand $E_{t+1,\omega}^{Agg,dep}$ of the EVs departure at the next hour and less than the maximum energy $E_{max,t,\omega}^{Agg}$ of the aggregator which is obtained by \eqref{DA_Smax}. In \eqref{DA_StLimit}, $\lambda$ is a percentage (e.g., 5\%) and a margin of $\lambda E_{max, t,\omega}^{Agg}$ in the lower and upper bounds is set up against the uncertainty of regulation signals. The constraint \eqref{DA_restS} indicates the rest of the aggregator's energy after the EVs depart at the next hour should not exceed the energy capacity of the rest EVs with considering the energy margin set up by $\lambda$. 
\begin{gather}
\begin{split}
E_{t,\omega}^{Agg} = &E_{t-1,\omega}^{Agg}-E_{t,\omega}^{Agg,dep}+E_{t,\omega}^{EV,arv}+ \\
& P_{t,\omega}^{Agg,ch} \times \eta_c - P_{t,\omega}^{Agg,dis} / \eta_d \label{DA_St}
\end{split} \\
E_{t,\omega}^{EV,dep} \leq E_{t,\omega}^{Agg,dep} \leq {N}_{t,\omega}^{EV,dep} \times E_{max}^{EV} \label{DA_Saggdep}\\
E_{t+1,\omega}^{Agg,dep}+\lambda E_{max,t,\omega}^{Agg} \leq E_{t,\omega}^{Agg} \leq (1-\lambda) E_{max,t,\omega}^{Agg} \label{DA_StLimit}\\
E_{max,t,\omega}^{Agg} = {N}_{t,\omega}^{EV} \times E_{max}^{EV} \label{DA_Smax}\\
E_{t,\omega}^{Agg}-E_{t+1,\omega}^{Agg,dep} \leq ({N}_{t,\omega}^{EV}-{N}_{t+1,\omega}^{EV,dep}) (1-\lambda) E_{max}^{EV} \label{DA_restS}
\end{gather}

According to the DA planning model \eqref{DA_DAobj}-\eqref{DA_restS}, the aggregator can estimate the revenue and determine the incentives for the next day. If the aggregator finds out there is a profitable opportunity to operate in the next day, it can activate the EV aggregation program and broadcast the incentives, which are obtained via the DA planning model, to the EV owners. \vspace{-0.5cm}
\subsection{Second Stage: Real-Time Operation of EV Aggregator}
%
After the aggregation program is activated and the optimal incentive $\pi$ has been delivered to the EV owners, the aggregator needs to participate in the RT energy and regulation markets for gaining revenues. Similar to \cite{do2018stochastic,vejdan2018maximizing}, the RT market is assumed to be an hourly cleared market in this paper and the market participants can submit the bids before the hour $t$ starts. In other words, the aggregator needs to keep bidding for the next operation hour in the RT market until the end of the aggregation program. To maximize the aggregator’s revenue in the RT market and generate the optimal bids for the next operation hour, a RT optimal bidding algorithm for the incentive-based EV aggregator is proposed. \par
Based on the DA planning model, a RT bidding model described in \eqref{RT_RTobj}-\eqref{RT_Edistau_LB} is proposed to maximize the profits for the rest of the operation hours $\hat{T}$ of the day. Since the incentive has already been determined via the DA planning model, the RT operation model only considers maximizing the credits from the energy and regulation markets as shown in \eqref{RT_RTobj}. In \eqref{RT_RTobj}-\eqref{RT_restS}, the ${LMP}'_t$, ${RMCCP}'_t$, ${RMPCP}'_t$, ${\rho}'_t$, ${\beta}'_t$, ${RegD}_t^{'up}$, and ${RegD}_t^{'down}$ are the market information predicted before the next operation hour of the RT markets. The other variables and parameters in \eqref{RT_RTobj}-\eqref{RT_restS} are set up similar to those in the DA planning model.
\begin{gather} 
Maximize \: (Credit'_E+Credit'_R) \label{RT_RTobj} 
\end{gather}
\begin{center}\vspace{-0.1cm} s.t. \eqref{RT_CreditE}-\eqref{RT_Edistau_LB}. \vspace{-0.1cm}\end{center}
\begin{gather} \allowdisplaybreaks
Credit'_{E} = \sum^{}_{t \in \hat{T}} (P_{t}^{E,dis}-P_{t}^{E,ch})\times LMP_t' \label{RT_CreditE}\\
\begin{split}
Credit'_{R} = & \sum^{}_{t \in \hat{T}} (P^R_{t}\times {{RMCCP}'_t} \times\ {{\rho}'_t} + \\ 
&P^R_{t}\times {{RMPCP}'_t} \times {\beta'_t} \times {{\rho}'_t}) \label{RT_CreditR}
\end{split} 
\end{gather}
\begin{gather} \allowdisplaybreaks
P_{max,t}^{Agg} = {N}_{t}^{EV} \times P_{max}^{EV} \\
0 \leq P_{t}^{E,ch} \leq P_{max,t}^{Agg}  \allowdisplaybreaks \\
0 \leq P_{t}^{E,dis} \leq P_{max,t}^{Agg} \allowdisplaybreaks\\
0 \leq P_{t}^{E,ch} \leq \delta_{t} \times M \allowdisplaybreaks\\
0 \leq P_{t}^{E,dis} \leq (1-\delta_{t}) \times M \allowdisplaybreaks\\
\delta_{t}  \in \{0,1\} \allowdisplaybreaks\\
0 \leq P_{t}^{E,ch}+P_{t}^{R} \leq P_{max,t}^{Agg} \allowdisplaybreaks\\
0 \leq P_{t}^{E,dis}+P_{t}^{R} \leq P_{max,t}^{Agg} \allowdisplaybreaks\\
P_{t}^{Agg,ch} = P_{t}^{E,ch}+P_{t}^R \times RegD_t^{'down} \allowdisplaybreaks\\
P_{t}^{Agg,dis} = P_{t}^{E,dis}+P_{t}^R \times RegD_t^{'up} \allowdisplaybreaks\\
\begin{split}
E_{t}^{Agg} = &E_{t-1}^{Agg}-E_{t}^{Agg,dep}+E_{t}^{EV,arv}+ \\
& P_{t}^{Agg,ch} \times \eta_c - P_{t}^{Agg,dis} / \eta_d 
\end{split} \\
E_{t}^{EV,dep} \leq E_{t}^{Agg,dep} \leq {N}_{t}^{EV,dep} \times E_{max}^{EV} \\
E_{t+1}^{Agg,dep}+\lambda E_{max,t}^{Agg} \leq E_{t}^{Agg} \leq (1-\lambda) E_{max,t}^{Agg} \label{RT_StLimit}\\
E_{max,t}^{Agg} = {N}_{t}^{EV} \times E_{max}^{EV} \\
E_{t}^{Agg}-E_{t+1}^{Agg,dep} \leq ({N}_{t}^{EV}-{N}_{t+1}^{EV,dep}) (1-\lambda)E_{max}^{EV} \label{RT_restS} \vspace{-0.5cm}
\end{gather} \par
\textbf{Additional Power Constraints}:
In addition to the power constraints similar to the DA planning model, there are four constraints \eqref{RT_Echtau_LB}-\eqref{RT_Edistau_LB} in the RT operation model for the power offered to the next operation hour $\tau$. During the RT operation, the aggregator needs to guarantee the EVs always leave the parking lot with their desired departure energy, which is important for the EV aggregator to keep operating the incentive-based aggregation program successfully: Having a satisfied customer. Therefore, the EV aggregator should schedule the charging/discharging power in the energy market to fulfill the energy demand of departure EVs against the uncertainty of regulation signals. \eqref{RT_Echtau_LB} indicates if the aggregator's energy at the beginning of hour $\tau$, which is obtained by $(E_{\tau-1}^{Agg}+E_{\tau}^{EV,arv}-E_{\tau}^{Agg,dep})$, is not enough for the desired departure energy $E_{\tau+1}^{EV,dep}$ at hour $\tau+1$, the aggregator must purchase a certain amount of energy from the energy market at hour $\tau$ to satisfy the energy demand of the departure EVs at hour $\tau+1$. Similarly, \eqref{RT_Edistau_UB} restricts the aggregator to over discharge in the energy market at hour $\tau$. As aforementioned, the $\lambda E_{max,\tau}^{Agg}$ is a margin for scheduling the aggregator's charging/discharging power, which is used against the uncertainty of regulation signals and improves the aggregator’s performance in the regulation market. \eqref{RT_Echtau_UB} and \eqref{RT_Edistau_LB} indicate the power offered at hour $\tau$ should be utilized to keep the aggregator from violating the $(1-\lambda)E_{max,\tau}^{Agg}$ at the end of hour $\tau$, which is also for better handling the uncertainty of the regulation signals at hour $\tau$ with an energy margin. \par
\begin{gather} \allowdisplaybreaks
\begin{split}
P_{\tau}^{E,ch} \times \eta_c \geq & \max\{0, \: (E_{\tau+1}^{EV,dep}+\lambda E_{max,\tau}^{Agg}) \\
                                   & -(E_{\tau-1}^{Agg}+E_{\tau}^{EV,arv}-E_{\tau}^{Agg,dep})\} \label{RT_Echtau_LB}
\end{split} \\
\begin{split}
P_{\tau}^{E,dis} / \eta_d \leq & \max\{0, \: (E_{\tau-1}^{Agg}+E_{\tau}^{EV,arv} -E_{\tau}^{Agg,dep})\\ 
                      & -(E_{\tau+1}^{EV,dep}+\lambda E_{max,\tau}^{Agg})\} \label{RT_Edistau_UB}
\end{split} 
\end{gather}
\begin{gather} \allowdisplaybreaks
\begin{split}
P_{\tau}^{E,ch} \times \eta_c \leq & \max\{0, \: (1-\lambda) E_{max,\tau}^{Agg}-(E_{\tau-1}^{Agg} \\
&+E_{\tau}^{EV,arv}-E_{\tau}^{Agg,dep}) \} \label{RT_Echtau_UB}
\end{split} \\
\begin{split}
P_{\tau}^{E,dis} / \eta_d \geq & \max\{0, \: (E_{\tau-1}^{Agg}+E_{\tau}^{EV,arv}-E_{\tau}^{Agg,dep}) \\
& -(1-\lambda)E_{max,\tau}^{Agg}\} \label{RT_Edistau_LB}
\end{split}
\end{gather} \par
At the end of hour $\tau$, the aggregator needs to calculate a provisional energy status $\hat{E}_{\tau}^{Agg}$ of the aggregator with the realized offers, including the $P_{\tau}^{E,ch}$, $P_{\tau}^{E,dis}$, and $P_{\tau}^R$, and the actual regulation signals represented by ${RegD}_{\tau}^{up}$ and ${RegD}_{\tau}^{down}$ as shown in \eqref{RT_ShatAgg}. It should be noted that the $E_{\tau-1}^{Agg}$ and $E_{\tau}^{Agg,dep}$ are obtained from the operation of hour $\tau-1$ and the process to obtain $E_{\tau-1}^{Agg}$ and $E_{\tau}^{Agg,dep}$ is introduced with Algorithm \ref{RT_Operation_Algorithm}, subsequently. \par
\begin{gather} \allowdisplaybreaks
\begin{split}
\hat{E}_{\tau}^{Agg} = & E_{\tau-1}^{Agg}-E_{\tau}^{Agg,dep}+E_{\tau}^{EV,arv} \\
					  & +(P_{\tau}^{E,ch}+P_{\tau}^R \times RegD_{\tau}^{down}) \eta_c \\
					  & -(P_{\tau}^{E,dis}+P_{\tau}^R \times RegD_{\tau}^{up})/ \eta_d \label{RT_ShatAgg}
\end{split}
\end{gather}
Due to the uncertainty of regulation signals, it is possible that the predicted regulation signals significantly deviate from the actual signals. Therefore, the $\hat{E}_{\tau}^{Agg}$  may violate the energy constraints at the end of hour $\tau$. In order to keep the aggregator operating within the energy requirements, three criteria corresponding to \eqref{RT_StLimit} and \eqref{RT_restS} are introduced in \eqref{RT_StLimit_Check1}-\eqref{RT_restS_Check1} for checking the $\hat{E}_{\tau}^{Agg}$. If $\hat{E}_{\tau}^{Agg}$ violates any one of these criteria, a post-process at the end of hour $\tau$ has to be taken to adjust the  $\hat{E}_{\tau}^{Agg}$ and $E_{\tau+1}^{Agg,dep}$.
\begin{gather} \allowdisplaybreaks
\hat{E}_{\tau}^{Agg} \geq E_{\tau+1}^{EV,dep} \label{RT_StLimit_Check1}\\
\hat{E}_{\tau}^{Agg} \leq E_{max,\tau}^{Agg} \label{RT_StLimit_Check2}\\
\hat{E}_{\tau}^{Agg}-E_{\tau+1}^{EV,dep} \leq ({N}_{\tau}^{EV}-{N}_{\tau+1}^{EV,dep}) \times E_{max}^{EV} \label{RT_restS_Check1}
\end{gather}
When $\hat{E}_{\tau}^{Agg} < E_{\tau+1}^{EV,dep}$, the energy difference $\Delta{E}_{\tau}^{Agg}$ obtained by \eqref{RT_deltaS_Check1} is utilized to calculate the nonperformance offer $\Delta{P_{\tau}^{R}}$, which is a shortfall performance of the $P_{\tau}^{R}$ in the regulation market, and the actual aggregator's energy $E_{\tau}^{Agg}$ after the adjustment via \eqref{RT_deltaPR} and \eqref{RT_deltaS_dep}, respectively. 
\begin{gather} \allowdisplaybreaks
\Delta{E_{\tau}^{Agg}} = \hat{E}_{\tau}^{Agg}-E_{\tau+1}^{EV,dep}  \label{RT_deltaS_Check1}\\
\Delta{P_{\tau}^{R}} = \Delta{E_{\tau}^{Agg}}/(RegD_{\tau}^{down}\eta_c-RegD_{\tau}^{up}/\eta_d) \label{RT_deltaPR}\\
E_{\tau}^{Agg} = \hat{E}_{\tau}^{Agg} - \Delta{E_{\tau}^{Agg}} \label{RT_deltaS_dep}
\end{gather}
If $\hat{E}_{\tau}^{Agg} > E_{max,\tau}^{Agg}$, the corresponding $\Delta{E}_{\tau}^{Agg}$, $\Delta{P_{\tau}^{R}}$, and $E_{\tau}^{Agg}$ are calculated by \eqref{RT_deltaS_Check2}, \eqref{RT_deltaPR} and \eqref{RT_deltaS_dep}, respectively.
\begin{gather}
\Delta{E_{\tau}^{Agg}} = \hat{E}_{\tau}^{Agg}-E_{max,\tau}^{Agg} \label{RT_deltaS_Check2}
\end{gather}
After the $E_{\tau}^{Agg}$ has been obtained, the $E_{\tau+1}^{Agg,dep}$ can be calculated by \eqref{RT_deltaSaggdep} and \eqref{RT_realSaggdep} if the criterion \eqref{RT_restS_Check1} is violated. 
\begin{gather} \allowdisplaybreaks
\begin{split}
\Delta{E_{\tau+1}^{Agg,dep}} = & E_{\tau}^{Agg}- E_{\tau+1}^{EV,dep} \\
&-({N}_{\tau}^{EV}-{N}_{\tau+1}^{EV,dep}) \times E_{max}^{EV} \label{RT_deltaSaggdep}\\
\end{split} \\
E_{\tau+1}^{Agg,dep} = E_{\tau+1}^{EV,dep}+\Delta{E_{\tau+1}^{Agg,dep}} \label{RT_realSaggdep}
\end{gather} \par
The process of the second stage of the proposed bidding algorithm is illustrated in Algorithm \ref{RT_Operation_Algorithm}. In the algorithm, $\tau$ indicates the next operation hour. Before the hour $\tau$, the aggregator needs to predict the market data for the rest operation hours $\hat{T}$ and solve the RT operation model in \eqref{RT_RTobj}-\eqref{RT_Edistau_LB} to obtain the bids for hour $\tau$. At the end of hour $\tau$, the  $\hat{E}_{\tau}^{Agg}$ is calculated by \eqref{RT_ShatAgg} and checked with the criteria \eqref{RT_StLimit_Check1}-\eqref{RT_restS_Check1}. Then, the obtained $E_{\tau}^{Agg}$ and $E_{\tau+1}^{Agg,dep}$ are used in the RT operation model of the next operation hour $\tau+1$. \par 
\begin{algorithm}
	\label{RT_Operation_Algorithm}
	\caption{Real-Time Operation} 
	\SetAlgoLined
	Initialization: $\tau=7$, $E_{\tau-1}^{Agg} = 0$, $E_{\tau}^{Agg,dep}=E_{\tau}^{EV,dep}$;
	\For {$\tau=7,8,\ldots,19$}{
		$\hat{T} = \{\tau,\ldots,19\}$\;
		Estimate market data for $t \in \hat{T}$ \;
		Solve the RT operation model \eqref{RT_RTobj}-\eqref{RT_Edistau_LB}\;
		Obtain $P_{\tau}^{E,ch}$, $P_{\tau}^{E,dis}$, and $P_{\tau}^{R}$\;
		Calculate $\hat{E}_{\tau}^{Agg}$ by \eqref{RT_ShatAgg}\;
		\uIf{$\hat{E}_{\tau}^{Agg} < E_{\tau+1}^{EV,dep}$}{
		 Obtain $\Delta{P_{\tau}^{R}}$ and $E_{\tau}^{Agg}$ by \eqref{RT_deltaS_Check1}, \eqref{RT_deltaPR}, \eqref{RT_deltaS_dep}
		}
		\uElseIf{$\hat{E}_{\tau}^{Agg} > E_{max,\tau}^{Agg}$}{
		 Obtain $\Delta{P_{\tau}^{R}}$ and $E_{\tau}^{Agg}$ by \eqref{RT_deltaS_Check2}, \eqref{RT_deltaPR}, \eqref{RT_deltaS_dep}
		}
		\Else{
		$\Delta{P_{\tau}^{R}} = 0$, $E_{\tau}^{Agg} = \hat{E}_{\tau}^{Agg}$\;
	    }
		\uIf{$E_{\tau}^{Agg}- E_{\tau+1}^{EV,dep} > ({N}_{\tau}^{EV}-{N}_{\tau+1}^{EV,dep}) \times E_{max}^{EV}$}{
		 Obtain $E_{\tau+1}^{Agg,dep}$ by \eqref{RT_deltaSaggdep}, \eqref{RT_realSaggdep}\;
		}
		\Else{
		$E_{\tau+1}^{Agg,dep} = E_{\tau+1}^{EV,dep}$\;
		}
	 Prepare $E_{\tau}^{Agg}$, $E_{\tau+1}^{Agg,dep}$ for the next hour $\tau+1$\;
	}
	\KwResult{$P_{t}^{E,ch}$, $P_{t}^{E,dis}$, $P^R_{t}$, $\Delta{P_{t}^{R}}$, $E_{t}^{Agg}$, and $E_{t+1}^{Agg,dep}$ for $t \in T$}
\end{algorithm} \setlength{\textfloatsep}{0pt}
It is worth pointing out that there will be no non-compliant issue and penalization for the aggregator in the energy market since the charging/discharging power cleared in the energy market has been enforced by \eqref{RT_Echtau_LB}-\eqref{RT_Edistau_LB} to satisfy the energy requirements of the coming operation hour. However, the predictions of the regulation signals can be significantly different from the actual values, the worst case for the aggregator is $\Delta{P}_{t}^R=P_{t}^R$, which means the aggregator is unable to perform any regulation services in the regulation market during the hour $t$. The departure energy required by the EVs can still be guaranteed under this circumstance. \par
As aforementioned, the performance score $\rho_{t}$ is used to penalize the market participants for failing to provide the cleared regulation capacity. In PJM, the performance score shown in \eqref{Definition_PerfromScore} is a combination of the precision score ($\rho_t^{precision}$), the correlation score ($\rho_t^{correlation}$), and the delay score ($\rho_t^{delay}$) ranging between 0 and 1 \cite{PJMPerformanceScore,xu2018optimal}. The precision score defined by \eqref{Definition_PrecisionScore} is used to evaluate the error of a regulation response.
\begin{gather}
\rho_t = (\rho_t^{precision}+\rho_t^{correlation}+\rho_t^{delay})/3 \label{Definition_PerfromScore}\\
\rho_t^{precision} = 1-(P_t^R-|\Delta{P}_t^R|)/P_t^R \label{Definition_PrecisionScore}
\end{gather}
The correlation score is implemented to measure the correlation between the regulation signals and resource's responses. The delay score is to evaluate the delay of the response between the regulation signal and the change of resource's output. Considering the outstanding ramping capability of the battery for following the regulation signal, both the correlation and delay scores can be approximated to be the same as the precision score at each hour for the simplification based on \cite{PJMPerformanceScore,xu2018optimal}. Therefore, $\rho_t$ is represented by $\rho_t^{precision}$. \par
Given the cleared offers, such as $P_t^{E,ch}$, $P_t^{E,dis}$, $P_t^R$, and the actual market data including $LMP_t$, $RMCCP_t$, $RMPCP_t$, $\beta_t$, and $\rho_t$, the credits of participation in the energy and regulation markets for the EV aggregator can be calculated based on \eqref{RT_CreditE} and \eqref{RT_CreditR}. \vspace{-0.3cm}
\vspace{-0.1cm}
\subsection{Prediction of Market Data}
\label{Sec_SARIMA}
\vspace{-0.1cm}
According to the studies for market data in \cite{zhao2017improving,donadee2014agc,macdonald2012demand}, the seasonality has been found in the LMP \cite{zhao2017improving}, regulation market clearing price \cite{macdonald2012demand}, and regulation signals \cite{donadee2014agc}. Meanwhile, a similar feature of the mileage ratio collected from PJM has been discovered, as well. Hence, to predict the market data in the DA planning and RT operation models, the seasonal ARIMA model ($ARIMA(p,d,q)\times {(P,D,Q)}_s$) \cite{box2015time} described by \eqref{eq_Seasonal_ARIMA} is implemented in this paper.
\begin{equation}\label{eq_Seasonal_ARIMA} 
{\phi}_p(B) {\Phi}_P(B^s) {\nabla}^d {\nabla}^D y_t = \mu +{\theta}_q(B) {\Theta}_Q(B^s) {\varepsilon}_t 
\end{equation} 
where $y_t$ is the time series fitted by the seasonal ARIMA model; $p$ is the non-seasonal auto-regression (AR) order; $d$ is the non-seasonal differencing order; $q$ is the non-seasonal moving-average (MA) order; $P$ is the seasonal AR order; $D$ is the seasonal differencing; $Q$ is the seasonal MA order; $S$ represents the time steps of repeating seasonal pattern; The error term ${\varepsilon}_t$ is an independent and identically distributed noise with zero mean and finite variance; $\mu $ is a constant term for the model offset; $B$ is the backward shift operator, i.e. $B^hy_t=y_{t-h}$. The operators $\phi_p(B)$, $\Phi_P(B^s)$,  $\nabla^d$, $\nabla^D$, $\theta_q(B)$, and $\Theta_Q(B^s)$ are described as \eqref{eq_Seasonal_ARIMA_2}-\eqref{eq_Seasonal_ARIMA_7}.
\begin{gather}
{\phi}_p(B) = 1 - {\phi }_1 B - {\phi}_2 B^2 -\dots - {\phi }_p B^p  \label{eq_Seasonal_ARIMA_2} \\
{\Phi}_P(B^s) = 1 - {\Phi}_1 B^s - {\Phi}_2 B^{2s} - \dots - {\Phi}_P B^{Ps} \label{eq_Seasonal_ARIMA_3} \\
{\nabla}^d = (1-B)^d \label{eq_Seasonal_ARIMA_4} \\
{\nabla}^D = (1-B^s)^D \label{eq_Seasonal_ARIMA_5} \\
{\theta}_q(B) = 1 - {\theta}_1 B - {\theta}_2 B^2 - \dots - {\theta}_q B^q \label{eq_Seasonal_ARIMA_6} \\
{\Theta }_Q(B^s) = 1-{\Theta}_1 B^s - {\Theta}_2 B^{2s} -\dots - {\Theta}_Q B^{Qs} \label{eq_Seasonal_ARIMA_7} 
\end{gather}
where ${\phi}_1 \dots {\phi}_p$, ${\Phi}_1 \dots {\Phi}_P$ , ${\theta}_1 \dots {\theta}_q$ and ${\Theta}_1 \dots {\Theta}_Q$ are the coefficients of the corresponding operators. \par
In order to fit the market data with the seasonal ARIMA model, a preprocess scheme is implemented in this paper. First, the market data is preprocessed by \eqref{eq_ARIMA_Pre1} to eliminate the outliers in data series:
\begin{equation} \label{eq_ARIMA_Pre1}
\overline{y}_{t} = \begin{cases}
{\mu}_{y}+3\sigma_y&,  if \; \; y_t > {\mu}_{y}+3\sigma_y  \\
y_t & ,  otherwise \\
{\mu}_{y}-3\sigma_y &, if \; \; y_t < {\mu}_{y}-3\sigma_y \\
\end{cases}
\end{equation}
where ${\mu}_{y}$ is the mean of $y_t$ and $\sigma_y$ is the standard deviation of $y_t$. Then, a natural logarithm transformation, given in \eqref{eq_ARIMA_Pre2}, is taken to reduce the fluctuations and make the series stationary for fitting the seasonal ARIMA model.
\begin{equation} \label{eq_ARIMA_Pre2}
\hat{y}_t = log(\overline{y}_{t}+c)
\end{equation}
where $\overline{y}_t$ is the series obtained by \eqref{eq_ARIMA_Pre1}; $c$ is a constant value to offset $\overline{y}_t$ for the logarithm transformation. The $\hat{y}_t$ acquired by \eqref{eq_ARIMA_Pre2} is utilized with the seasonal ARIMA model for the prediction. \vspace{-0.3cm}

\section{Model Performance} \label{model_performance}

\subsection{Incentives and EV Responses}
In order to study the performance of the proposed models and algorithms, 200 EVs are assumed to participate in the EV aggregation program. The power and capacity of each EV are set to 50 kW and 50 kWh \cite{EVpowercapacity} in this paper. Hence, the maximum power and energy capacity of the aggregator are 10 MW and 10 MWh, respectively. Based on the revenue analysis for energy storage systems in energy and regulation markets \cite{zhao2020revenue,byrne2016estimating}, the average daily revenue for a 10 MW/ 10 MWh energy storage system with considering the battery degradation and perfect price forecasting is about \$6,300/day. Since the EV aggregator proposed in this paper is operating between 7:00 and 19:00, the maximum credit obtained from energy and regulation markets during the 13 operating hours is expected to be about \$3,500/day. Accordingly, the maximum reward for the EV owners is set to be \$2500/day, including the fixed reward $\hat{\pi}$ of \$1000/day and the incentive $\pi$ up to \$1500/day. The credit less the rewards is considered as the EV aggregator’s gross revenue, covering the operating cost. \par
\begin{figure}[H]
	\centering
	\includegraphics[width=2.4 in]{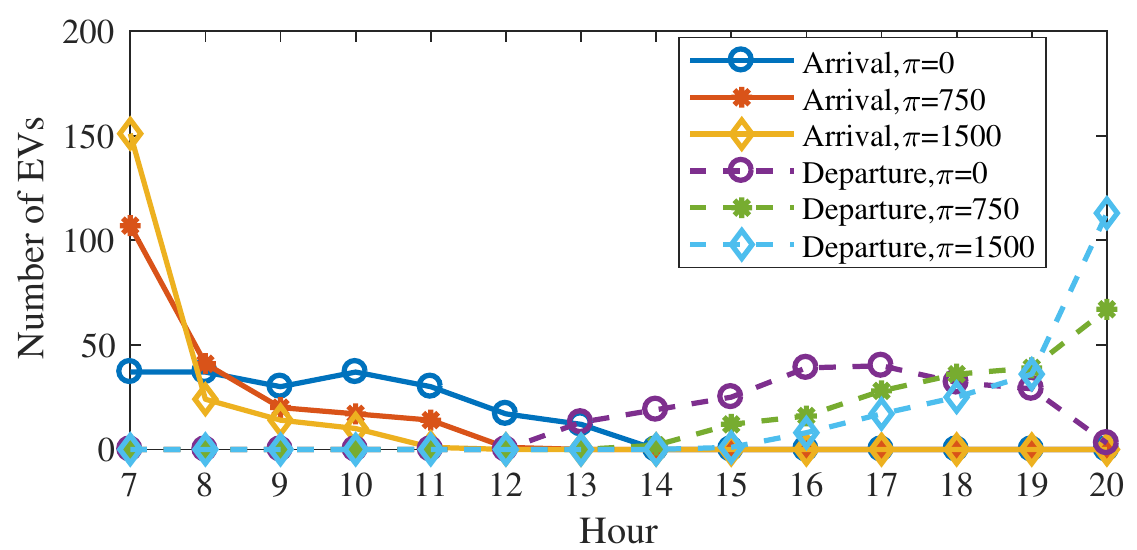}
	\caption{Arrival and departure EVs with different incentives.}
	\label{Response_Arv_Dep_Nev} \vspace{-0.5cm}
\end{figure}

The initial behaviors of each EV are sampled via the truncated Gaussian functions introduced in Section \ref{Sec_EV_Behaviors}. Given the initial behaviors of each EV, the response step functions of the behaviors, including $t_{arv}^i$, $t_{dep}^i$, $SOC_{arv}^i$, and $SOC_{dep}^i$, can be generated with the incentive $\pi$ ranging from 0 to 1500. Consequently, the step functions of $N_t^{EV,arv}$, $ N_t^{EV,dep}$, $E_t^{EV,arv}$, $E_t^{EV,dep}$ for the whole EV fleet corresponding to different incentive levels can be derived. As a reference, the $N_t^{EV,arv}$ and $N_t^{EV,dep}$ of $\pi=$ 0, 750, and 1500 are plotted in Fig. \ref{Response_Arv_Dep_Nev}, where the horizontal axis indicates the operating hours and the vertical axis shows the number of arrival/departure EVs. It can be clearly seen that EVs tend to arrive earlier and depart later when the incentive $\pi$ increase from 0 to 1500. \vspace{-0.5cm}
 \par
\subsection{First Stage: DA Planning}
The real market data, such as LMP, RMCCP, RMPCP, mileage ratio and regulation signals, between 1/1/2018 and 12/31/2018 are collected from PJM Data Miner 2 \cite{PJMDataMiner2} to test the performance of the proposed models in the energy and regulation markets. To mitigate the impact of the transmission congestion on influencing the analysis of model performance, the LMP of PJM-RTO, which is an aggregated LMP of the whole energy market, is utilized in the following study to analyze the overall performance of the proposed model in the markets first. Then, the proposed two-stage optimal bidding algorithm is evaluated with the LMPs of different cities located in PJM.\par
\begin{figure}[H]
	\vspace{-0.5cm}
	\centering
	\includegraphics[width=2.5 in]{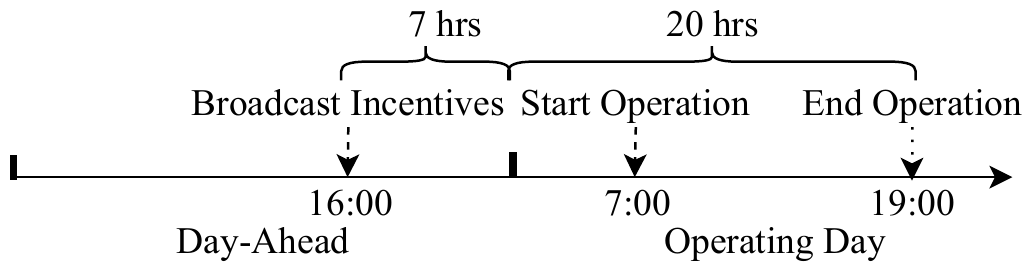}
	\caption{Timeline of DA planning and RT operation.}
	\label{DA_RT_Timeline} \vspace{-0.4cm}
\end{figure}

In this paper, the EV aggregator is designed to broadcast the incentive information to the EV owners at 16:00 before the operating day, as shown in Fig. \ref{DA_RT_Timeline}, in order to obtain more information on the RT market and let the EV owners get the incentive information before most of them leave the parking lot. Since the aggregation program starts at 7:00 on an operating day, the EV aggregator needs to predict the market data for the next 27 hours, including the rest 7 hours of the DA day and the 20 hours of the operating day before the operation ends. 
 \par
The introduced forecasting method based on the seasonal ARIMA model has been implemented to predict the market data. According to \cite{zhao2017improving,box2015time}, the seasonality of the market data is found to be 24 via the analysis of autocorrelation function (ACF) and partial autocorrelation function (PACF). Meanwhile, the other orders of the seasonal ARIMA model are determined with the values listed in Table \ref{tb_ARIMA_Orders} depending on the stationarity of the forecasting time series and the analysis of ACF and PACF. In this paper, the market data before 2/26/2018 (Monday) are used to train the corresponding seasonal ARIMA models. Therefore, the proposed models are tested during the period between 2/26/2018 (Monday) and 12/30/2018 (Sunday). \vspace{-0.5cm}
\begin{table}[H]
	\centering
	\caption{Orders of Seasonal ARIMA Model}\label{tb_ARIMA_Orders}
	\begin{tabular}{c|c|c|c|c|c|c|c} \hline 
	    \textbf{Orders} & \textbf{p} & \textbf{d} & \textbf{q} & \textbf{P} & \textbf{D} & \textbf{Q} & \textbf{s}\\ \hline 
	    \textbf{Values} & 2, 3 & 0, 1 & 0, 1 & 1 & 0, 1 & 0, 1 & 24\\ \hline 
	\end{tabular} \vspace{-0.4cm}
\end{table} \par
After the DA planning model is solved with the forecasting market data, the estimated revenue for the operating day can be calculated. Since the intention of the proposed EV aggregator at the parking lot is to benefit the EV owners with the incentives and conduct more regulation resources to the power system, the aggregation program is set to be activated whenever the estimated revenue is larger than \$0. Meanwhile, it should be mentioned that the EV aggregation program is activated only when the operating day is not a weekend or holiday. Between 2/26/2018 and 12/30/2018, there are 213 working days valid for the EV aggregator at the parking lot to participate in the markets. According to the estimated revenue obtained via the DA planning model, the EV aggregation program is activated for 125 out of 213 working days. During these days, the EV owners can be rewarded with \$229,625, which means the 200 EV owners can share \$1,837/day and each EV owner is expected to get a reward of \$9.19/day for participating in the EV aggregation program. \vspace{-0.5cm}
\subsection{Second Stage: RT Operation}
\begin{figure}[H]
	\vspace{-0.4cm}
	\centering
	\includegraphics[width=2.5 in]{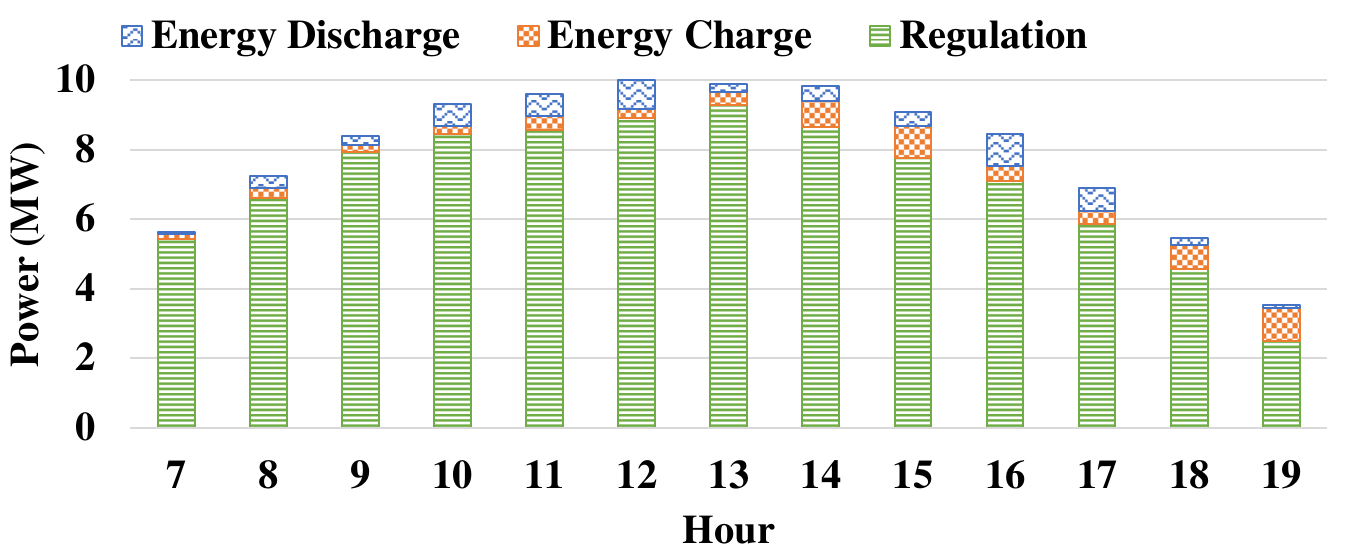}
	\caption{Average offers in energy and regulation markets by operating hours.}
	\label{Hourly_Average_Offers} \vspace{-0.4cm}
\end{figure}

\vspace{-0.1cm}
When the operating days and the corresponding incentives have been determined by the DA planning model, the EV aggregator needs to participate in the RT energy and regulation markets for the credits. According to the designed RT operation algorithm, the hourly offers, including $P_t^{E,ch}$, $P_t^{E,dis}$, and $P_t^R$, and the performance score can be obtained. In Fig. \ref{Hourly_Average_Offers}, the average offers by the operating hours are plotted. As shown in Fig. \ref{Hourly_Average_Offers}, the aggregator offers most of its capacity to the regulation market, which indicates the operation model tends to obtain the credits by providing regulation service instead of energy arbitrage. The aggregator keeps bidding a certain amount of power charging in the energy market after hour 13, as the EVs start leaving at hour 13, and the aggregator needs to guarantee that the EVs leave with their desired energy level.
 \par
\begin{figure}[H]
	\vspace{-0.5cm}
	\centering
	\includegraphics[width=2.5 in]{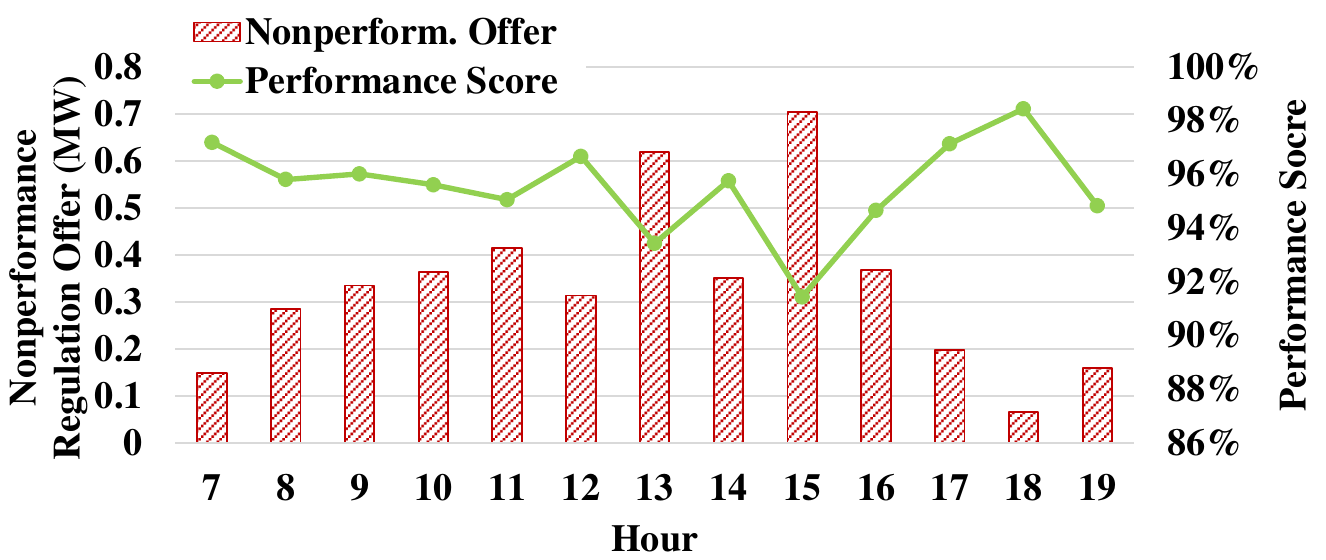}\vspace{-0.1cm}
	\caption{Hourly average nonperformance regulation offer and precision score.}
	\label{Norperform_Score} \vspace{-0.4cm}
\end{figure}

As aforesaid, the aggregator needs to adjust its actual output in the regulation market for complying with the energy constraints, especially when the actual regulation signals are quite different from the forecasting regulation signals. Therefore, the aggregator may fail to perform the committed regulation offer and get a penalty evaluated by the performance score for some hours. In Fig. \ref{Norperform_Score}, the average nonperformance power and the precision score at different operating hours are plotted. The left and right vertical indices in Fig. \ref{Norperform_Score} indicate the MW of nonperformance offer and the performance score, respectively. As shown in Fig. \ref{Norperform_Score}, the performance scores at all operating hours are over 91\% and the average performance score of the aggregator is over 95\%, which means the aggregator performs well on complying with the cleared regulation offers. Since the aggregator bids most of its capacity in the regulation market as displayed in Fig. \ref{Hourly_Average_Offers}, it is very important that the aggregator's profit is not affected by a poor performance in the regulation market.
 \par
The total credit of the aggregator obtained from the energy and regulation markets during the 125 operating days is \$360,833.73 and the average daily credit is \$2,886.67. After issuing the incentives to EV owners, the aggregator's revenue is \$131,208.73.
With a further analysis of the market credits, the average daily credits of energy and regulation markets are \$46.41/day and \$2,840.26/day, respectively. The credit from the energy market is only 1.63\% of the regulation market. In other words, the energy market mainly helps the aggregator to comply with the energy requirements instead of providing a meaningful revenue. For comparison, a base case for participating in RT energy and regulation markets is included, where the aggregator maximizes its credits without using the incentives to change the EVs' behaviors or implementing the energy margin $\lambda$ in the optimization models against the uncertainty of the regulation signals. It should be noted that the EV aggregation program without the procedure of DA planning is activated everyday. With this strategy, the EV aggregator can only obtain \$1,456.66 per day with a performance score of 85.9\%.\par
%
%
In addition to the 125 operating days determined via the DA planning model, the proposed models and algorithm have also been tested for the rest 88 working days in the test period. The average daily credit obtained from the RT markets is \$1,415.89, which is only 49\% the average credit earned during the operating days. In other words, the profit margin would get shrunk if the aggregator kept activating the incentive-based EV aggregation program on those working days. It also means the aggregator could suffer a higher risk of loss without the proposed DA planning model by still providing blind incentives. \par
%
%
After the overall performances of the proposed algorithm have been evaluated with the LMP of PJM-RTO, the LMPs of the price nodes located in several large cities \cite{PJMLMPZipcode}, including: Chicago, Cleveland, Pittsburgh, and Philadelphia, are selected for a further analysis of the bidding algorithm for the EV aggregator in workplace parking lots. As shown in Table \ref{tb_Algorithm_Performance_Cities}, the average daily credits earned by the EV aggregation program in different cities are close to the credit of the overall performances, which indicates the proposed incentive-based EV aggregation program is capable of benefitting the EV owners and aggregators at different locations.

%

\begin{table}[H]
	\centering
	\vspace{-0.5cm}
	\caption{Average Daily Performances in Different Cases}\label{tb_Algorithm_Performance_Cities}
	\begin{tabular}{c|c|c|c|c} \hline 
		\textbf{Case} & \textbf{Avg. Credit} & \textbf{EV} & \textbf{Aggregator} & \textbf{Avg. \boldmath$\rho_t$} \\ \hline 
		Base & \$1,456.66 & \$1,000.00 & \$456.66 & 85.9\% \\ 
		PJM-RTO & \$2,886.67 & \$1,837.00 & \$1,049.67 & 95.6\% \\ 
		Chicago, IL & \$2,869.32 & \$1,820.89 & \$1,048.43 & 95.7\% \\ 
		Cleveland, OH & \$2,885.81 & \$1,838.08 & \$1,047.73 & 95.1\% \\  
		Pittsburgh, PA & \$2,879.95 & \$1,838.67 & \$1,041.28 & 95.6\% \\  
		Philadelphia, PA & \$2,943.88 & \$1,812.02 & \$1,131.86 & 96.2\% \\ \hline 
	\end{tabular} \vspace{-0.3cm}
\end{table}

\vspace{-0.5cm}
\section{Conclusion} \label{conclusion}
\vspace{-0.1cm}
A two-stage optimal bidding algorithm for an incentive-based aggregator of EVs at workplace parking lots has been proposed in this paper. Based on the features and incentive responses of individual vehicles’ behaviors, such as arrival/departure times, arrival energy, and desired departure energy, the behaviors of the whole EV fleet aggregator have been characterized. A DA planning model considering the incentive response functions has been proposed to determine whether or not to provide incentives and the incentive level for the aggregator. With a specified incentive, a RT operation algorithm has been designed for maximizing the aggregator’s profit in the RT energy and regulation markets while satisfying the EVs’ energy demand. The proposed models have been tested using the real PJM energy and regulation market data in 2018. The results show that 200 EV owners are expected to get an average daily reward of \$1,837 in total, and the aggregator can have a profit of \$1,049.67 per day. The revenue analysis of the overall performances has shown the proposed two-stage optimal bidding algorithm is effective in benefitting both the EV owners and the aggregator for participating in the energy and regulation markets. The revenue analysis has also shown that the incentive-based EV aggregator at workplace parking lots is capable of having a stable performance when participating in energy and regulation markets at different locations, such as the metropolitan areas. \vspace{-0.3cm}
%


%






%
\bibliographystyle{IEEEtran}
\bibliography{references}

\begin{thebibliography}{10}
\providecommand{\url}[1]{#1}
\csname url@samestyle\endcsname
\providecommand{\newblock}{\relax}
\providecommand{\bibinfo}[2]{#2}
\providecommand{\BIBentrySTDinterwordspacing}{\spaceskip=0pt\relax}
\providecommand{\BIBentryALTinterwordstretchfactor}{4}
\providecommand{\BIBentryALTinterwordspacing}{\spaceskip=\fontdimen2\font plus
\BIBentryALTinterwordstretchfactor\fontdimen3\font minus
  \fontdimen4\font\relax}
\providecommand{\BIBforeignlanguage}[2]{{%
\expandafter\ifx\csname l@#1\endcsname\relax
\typeout{** WARNING: IEEEtran.bst: No hyphenation pattern has been}%
\typeout{** loaded for the language `#1'. Using the pattern for}%
\typeout{** the default language instead.}%
\else
\language=\csname l@#1\endcsname
\fi
#2}}
\providecommand{\BIBdecl}{\relax}
\BIBdecl

\bibitem{PJMEVtest}
W.~Kempton, V.~Udo, K.~Huber, K.~Komara, S.~Letendre, S.~Baker, D.~Brunner, and
  N.~Pearre, ``{A test of vehicle-to-grid (V2G) for energy storage and
  frequency regulation in the PJM system},'' \emph{Results from an
  Industry-University Research Partnership}, vol.~32, 2008.

\bibitem{ortega2012electric}
M.~A. Ortega-Vazquez, F.~Bouffard, and V.~Silva, ``{Electric vehicle
  aggregator/system operator coordination for charging scheduling and services
  procurement},'' \emph{IEEE Trans. Power Syst.}, vol.~28, no.~2, pp.
  1806--1815, 2012.

\bibitem{bessa2010role}
R.~J. Bessa and M.~A. Matos, ``{The role of an aggregator agent for EV in the
  electricity market},'' 2010.

\bibitem{2018EVVolumes}
``{Global EV Sales for 2018},''
  \url{http://www.ev-volumes.com/country/total-world-plug-in-vehicle-volumes/},
  accessed: 2020-05-20.

\bibitem{seddig2019two}
K.~Seddig, P.~Jochem, and W.~Fichtner, ``{Two-stage stochastic optimization for
  cost-minimal charging of electric vehicles at public charging stations with
  photovoltaics},'' \emph{Appl. energy}, vol. 242, pp. 769--781, 2019.

\bibitem{neyestani2014allocation}
N.~Neyestani, M.~Y. Damavandi, M.~Shafie-Khah, J.~Contreras, and J.~P. Catalao,
  ``{Allocation of plug-in vehicles' parking lots in distribution systems
  considering network-constrained objectives},'' \emph{IEEE Trans. Power
  Syst.}, vol.~30, no.~5, pp. 2643--2656, 2014.

\bibitem{zhao2017improving}
Z.~Zhao, C.~Wang, M.~Nokleby, and C.~J. Miller, ``{Improving short-term
  electricity price forecasting using day-ahead LMP with ARIMA models},'' in
  \emph{2017 IEEE PES General Meeting}.\hskip 1em plus 0.5em minus 0.4em\relax
  IEEE, 2017, pp. 1--5.

\bibitem{zhao2020revenue}
Z.~Zhao, C.~Wang, and M.~H. Nazari, ``{Revenue Analysis and Optimal Placement
  of Stationary and Transportable Energy Storage Systems in Energy and
  Frequency Regulation Markets},'' \emph{arXiv preprint arXiv:2001.01771},
  2020.

\bibitem{donadee2014agc}
J.~Donadee and J.~Wang, ``{AGC signal modeling for energy storage
  operations},'' \emph{IEEE Trans. Power Syst.}, vol.~29, no.~5, pp.
  2567--2568, 2014.

\bibitem{shafie2016optimal}
M.~Shafie-khah, E.~Heydarian-Forushani, G.~J. Os{\'o}rio, F.~A. Gil, J.~Aghaei,
  M.~Barani, and J.~P. Catal{\~a}o, ``{Optimal behavior of electric vehicle
  parking lots as demand response aggregation agents},'' \emph{IEEE Trans.
  Smart Grid}, vol.~7, no.~6, pp. 2654--2665, 2016.

\bibitem{nguyen2018bidding}
H.~T. Nguyen, L.~B. Le, and Z.~Wang, ``A bidding strategy for virtual power
  plants with the intraday demand response exchange market using the stochastic
  programming,'' \emph{IEEE Trans. Ind. Appl.}, vol.~54, no.~4, pp. 3044--3055,
  2018.

\bibitem{wang2016two}
Z.~Wang and Y.~He, ``Two-stage optimal demand response with battery energy
  storage systems,'' \emph{IET Gener. Transm. Distrib.}, vol.~10, no.~5, pp.
  1286--1293, 2016.

\bibitem{zhong2013coupon}
H.~Zhong, L.~Xie, and Q.~Xia, ``{Coupon incentive-based demand response: Theory
  and case study},'' \emph{IEEE Trans. Power Syst.}, vol.~28, no.~2, pp.
  1266--1276, 2013.

\bibitem{fang2016coupon}
X.~Fang, Q.~Hu, F.~Li, B.~Wang, and Y.~Li, ``{Coupon-based demand response
  considering wind power uncertainty: A strategic bidding model for load
  serving entities},'' \emph{IEEE Trans. Power Syst.}, vol.~31, no.~2, pp.
  1025--1037, 2016.

\bibitem{do2018stochastic}
J.~C. do~Prado and W.~Qiao, ``{A stochastic decision-making model for an
  electricity retailer with intermittent renewable energy and short-term demand
  response},'' \emph{IEEE Trans. Smart Grid}, vol.~10, no.~3, pp. 2581--2592,
  2018.

\bibitem{sarker2014optimal}
M.~R. Sarker, M.~A. Ortega-Vazquez, and D.~S. Kirschen, ``{Optimal coordination
  and scheduling of demand response via monetary incentives},'' \emph{IEEE
  Trans. Smart Grid}, vol.~6, no.~3, pp. 1341--1352, 2014.

\bibitem{byrne2016estimating}
R.~H. Byrne, R.~J. Concepcion, and C.~A. Silva-Monroy, ``{Estimating potential
  revenue from electrical energy storage in PJM},'' in \emph{2016 IEEE PES
  General Meeting}.\hskip 1em plus 0.5em minus 0.4em\relax IEEE, 2016, pp.
  1--5.

\bibitem{bessa2012optimized}
R.~J. Bessa, M.~A. Matos, F.~J. Soares, and J.~A.~P. Lopes, ``{Optimized
  bidding of a EV aggregation agent in the electricity market},'' \emph{IEEE
  Trans. Smart Grid}, vol.~3, no.~1, pp. 443--452, 2012.

\bibitem{vagropoulos2013optimal}
S.~I. Vagropoulos and A.~G. Bakirtzis, ``{Optimal bidding strategy for electric
  vehicle aggregators in electricity markets},'' \emph{IEEE Trans. Power
  Syst.}, vol.~28, no.~4, pp. 4031--4041, 2013.

\bibitem{sarker2015optimal}
M.~R. Sarker, Y.~Dvorkin, and M.~A. Ortega-Vazquez, ``{Optimal participation of
  an electric vehicle aggregator in day-ahead energy and reserve markets},''
  \emph{IEEE Trans. Power Syst.}, vol.~31, no.~5, pp. 3506--3515, 2015.

\bibitem{yao2016optimization}
E.~Yao, V.~W. Wong, and R.~Schober, ``{Optimization of aggregate capacity of
  PEVs for frequency regulation service in day-ahead market},'' \emph{IEEE
  Trans. Smart Grid}, vol.~9, no.~4, pp. 3519--3529, 2016.

\bibitem{liu2018two}
Z.~Liu, Q.~Wu, K.~Ma, M.~Shahidehpour, Y.~Xue, and S.~Huang, ``{Two-stage
  optimal scheduling of electric vehicle charging based on transactive
  control},'' \emph{IEEE Trans. Smart Grid}, vol.~10, no.~3, pp. 2948--2958,
  2018.

\bibitem{vaya2014optimal}
M.~G. Vay{\'a} and G.~Andersson, ``{Optimal bidding strategy of a plug-in
  electric vehicle aggregator in day-ahead electricity markets under
  uncertainty},'' \emph{IEEE Trans. Power Syst.}, vol.~30, no.~5, pp.
  2375--2385, 2014.

\bibitem{PJMMaunual11}
``{PJM Manual 11: Energy and Ancillary Services Market Operations},''
  \url{https://www.pjm.com/~/media/documents/manuals/m11.ashx}, accessed:
  2020-05-20.

\bibitem{xu2018optimal}
B.~Xu, Y.~Shi, D.~S. Kirschen, and B.~Zhang, ``{Optimal battery participation
  in frequency regulation markets},'' \emph{IEEE Trans. Power Syst.}, vol.~33,
  no.~6, pp. 6715--6725, 2018.

\bibitem{vejdan2018maximizing}
S.~Vejdan and S.~Grijalva, ``{Maximizing the revenue of energy storage
  participants in day-ahead and real-time markets},'' in \emph{2018 Clemson
  University Power Systems Conference}.\hskip 1em plus 0.5em minus 0.4em\relax
  IEEE, 2018, pp. 1--6.

\bibitem{PJMPerformanceScore}
``{PJM Performance Scoring},''
  \url{https://www.pjm.com/-/media/committees-groups/task-forces/rmistf/postings/performance-scoring-design-component.ashx?la=en},
  accessed: 2020-05-20.

\bibitem{macdonald2012demand}
J.~MacDonald, P.~Cappers, D.~Callaway, and S.~Kiliccote, ``{Demand response
  providing ancillary services: A comparison of opportunities and challenges in
  the US wholesale markets},'' \emph{Grid Interop}, 2012.

\bibitem{box2015time}
G.~E. Box, G.~M. Jenkins, G.~C. Reinsel, and G.~M. Ljung, \emph{{Time series
  analysis: forecasting and control}}.\hskip 1em plus 0.5em minus 0.4em\relax
  John Wiley \& Sons, 2015.

\bibitem{EVpowercapacity}
``{Charging Station},'' \url{https://en.wikipedia.org/wiki/Charging_station},
  accessed: 2020-05-20.

\bibitem{PJMDataMiner2}
``{PJM Data Miner 2},'' \url{https://dataminer2.pjm.com/list}, accessed:
  2020-05-20.

\bibitem{PJMLMPZipcode}
``{PJM LMP Model Information},''
  \url{https://www.pjm.com/markets-and-operations/energy/lmp-model-info.aspx},
  accessed: 2020-05-20.

\end{thebibliography}

%








\end{document}